\documentclass[fp,twocolumn]{jpsj3}
\usepackage{txfonts}
\usepackage{graphicx}
\usepackage{amsmath}
\usepackage{amssymb}
\usepackage{amsfonts}

\def\rs{{\rm s}}

\title{Translational-Symmetry-Broken Magnetization Plateaux of the $S=3/2$ Anisotropic Antiferromagnetic Chain}

\author{Tomohide Kawatsu$^{1}$
, Haruto Suzuki$^1$, Masaru Hashimoto$^1$
, Koki Doi$^1$, Tomoki Houda$^1$, Rito Furuchi$^1$, Hiroki Nakano$^1$, 
Kiyomi~Okamoto$^1$, and T\^oru Sakai$^{1,2}$\thanks{sakai@spring8.or.jp} }
\inst{$^1$Graduate School of Science, University of Hyogo, 3-2-1 Kouto, Kamigori, Ako-gun, Hyogo 678-1297, Japan \\
$^2$National Institute for Quantum Science and Technology (QST) SPring-8, Kouto 1-1-1, Sayo, Sayo-gun, Hyogo 679-5148 Japan \\
} 

\abst{The magnetization process of the $S=3/2$ quantum spin chain with the $XXZ$ anisotropy and
the single-ion anisotropy $D$ is investigated 
using the numerical diagonalization of finite-size clusters and the level spectroscopy analysis. 
We obtain the phase diagrams at 1/3 and 2/3 of the saturation magnetization
to find that the translational-symmetry-broken magnetization plateau appears
for the first time.
The similarity and the difference between the phase diagrams 
of the present model and the related models are discussed by use of
the discrete parameters of the models. 
In addition several typical magnetization curves are presented.  }

\begin{document}
\maketitle

\section{Introduction}

The magnetization plateau is one of interesting topics in the field of the condensed matter physics. 
For the one-dimensional case,
it was proposed as the Haldane gap which appears in the magnetization process\cite{oshikawa}. 
Based on the Lieb-Schultz-Mattis theorem\cite{LSM}, the rigorous necessary condition for the 
appearance of the magnetization plateau in the one-dimensional quantum spin systems 
was derived as the following form\cite{oshikawa}
\begin{eqnarray}
   Q(S-\tilde m)=n,~~~~~n \in {\mathbb N}
\label{condition}
\end{eqnarray}
where $S$ and $\tilde m$ are the total spin and the magnetization per unit cell, respectively, 
$Q$ is the periodicity of the ground sate,
and $n$ is a positive integer. 
Several magnetization plateaux with $Q=1$ have been theoretically predicted using 
some numerical analyses\cite{sakai1,
kitazawa-okamoto,
honecker1,okamoto1,gu-su,honecker2,
ananikian,ueno,filho,hida,okamoto-ssc,oka-kita,gong2,liu2,gong,mahdavifar,jiang,liu3,sugimoto3,
sugimoto1,sasaki,rahaman,cabra,okamoto-tube,li,alecio,farchakh,
dey,yamamoto,
sakai2,tonegawa,tenorio,liu,karlova,cabra2,chen},
 and experimentally observed\cite{kikuchi,morita,yamaguchi,yin}. 
 The translational symmetry broken magnetization plateaux with $Q=2$ were also 
 theoretically predicted in several systems\cite{totsuka,
okunishi1,
okunishi2,
metavitsiadis,
nakano,okazaki1,okazaki2,nakasu,sakai3,okamoto2,okamoto3,michaud,kohshiro},  
 based on the mechanism of the spontaneous dimer formation caused by the spin frustration. 

Recently we investigated the magnetization plateaux at half of the saturation magnetization
of $S=1$\cite{sakai4}, $S=2$\cite{sakai2019,yamada} antiferromagnetic chains 
with the exchange anisotropy $\lambda$ and the single-ion one $D$,
described by
\begin{eqnarray}
{\cal H}&=&{\cal H}_0+ {\cal H}_Z, \label{ham}
\\
   {\cal H}_0 
  &=& \sum _{j=1}^L (S_j^x S_{j+1}^x + S_j^y S_{j+1}^y + \lambda S_j^z S_{j+1}^z)
     +D\sum_{j=1}^L (S_j^z)^2, \nonumber \\
{\cal H}_Z&=&-H\sum_{j=1}^{L}S_j^z,
\nonumber
\end{eqnarray}
where $H$ is the external magnetic field along the $z$-direction.
Here, for convenience,
we define the relative magnetization $m$ as 
\begin{equation}
   m = {M \over M_{\rm s}},~~~~~
   M \equiv \sum_{j=1}^L S_j^z,~~~~~
   M_{\rm s} = LS,
   \label{eq:rel-mag}
\end{equation}
where $M$ is the magnetization
and $M_{\rm s}$ is the saturation magnetization.

The phase diagrams of the above models with $S=1$ and $S=2$ at $m=1/2$
are quite different from each other.
Namely, there appeared the no-plateau phase and the N\'eel plateau phase ($Q=2$) for the
$S=1$ case,\cite{sakai4}
whereas, in addition to those,  
the Haldane plateau ($Q=1$) and the large-$D$ plateau ($Q=1$) phases appeared
for the $S=2$ case.\cite{sakai2019,yamada}
We note that, $Q=2$ is necessary for the plateau in case of $S=1$,
whereas $Q=1$ is sufficient in case of $S=2$.
At zero magnetization $m=0$,
the phase diagrams of the above model with $S=1$\cite{den-nijs,chen-S=1} and
$S=2$\cite{tone-S=2,kjall}
are also rather different from each other,
although $Q=1$ is sufficient for the plateau (often called spin gap for the $m=0$ case)
in both cases.

%
%
%
In this paper, considering the above situation,
we investigate the magnetization plateau of the model (\ref{ham}) with $S=3/2$
at $m=1/3$ and $2/3$.
For the realization of the plateau,
$Q=1$ is sufficient for $m=1/3$,
whereas $Q=2$ is necessary for $m=2/3$.
Thus the comparison of the phase diagrams of this model with $m=1/3$ and $2/3$
with those $m=1/2$ ones with $S=1$ and $S=2$ is an interesting problem.
We use the numerical diagonalization of finite-size clusters and 
the level spectroscopy analysis. 
For the $m=1/3$ case,
although the phase diagram of limited region was obtained,\cite{okamoto1}
no N\'eel plateau has been found so far.
As far as we know,
there has been no report on the 2/3 magnetization plateau.

We will present an extended phase diagram at $m=1/3$ and 
also that at $m=2/3$ for the first time. 
In addition the magnetization curves for several typical parameters 
will be presented.

\section{Model and numerical calculation}

We investigate the magnetization process of the $S=3/2$ antiferromagnetic chain 
with the exchange anisotropy $\lambda$ and the single-ion one $D$ described by 
(\ref{ham}).
We consider the case when the coupling anisotropy is of easy-axis ($\lambda > 1$) 
and the single-ion one is of easy-plane ($D>0$). 
Then they compete with each other. 

In order to consider the possibility of the magnetization plateau, 
we calculate the lowest energy eigenvalue in the subspace of $M$, 
which is denoted as $E(L,M)$, using the Lanczos algorithm. 
The system size $L$ is up to 12, and the periodic boundary condition 
is applied. 
Only when the phase boundary of the $Q=1$ magnetization plateau at $m=1/3$ is considered 
in the next section, 
we also use the twisted boundary condition, 
namely the signs of $S_1^x$ and $S_1^y$ are changed on the connection 
of the sites $L$ and 1.

\section{Magnetization plateaux}

We consider the magnetization plateaux at $m=1/3$ and $m=2/3$ and 
obtain the phase diagrams with respect to the anisotropies $\lambda$ and $D$ 
at each magnetization in this section. 

\subsection{$m=1/3$}

\begin{figure}[ht]
\centerline{\includegraphics[width=0.85\linewidth,angle=0]{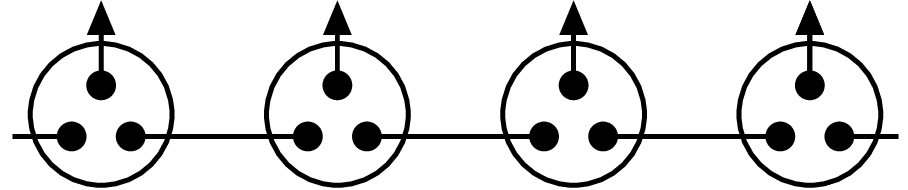}}%
\caption{\label{haldane} 
Schematic picture of the Haldane mechanism of the $m=1/3$ plateau.
A big open circle represents $S=3/2$ spin composed of three $S=1/2$ component spins
denoted by small dots.
Component spin a same big circle couple ferromagnetically with each other,
while those in different big circles couple antiferromagnetically with each other.
An up-arrowed dot is in the $S^z=1/2$ state.
Two dots connected by a thin line form a singlet pair $(1/\sqrt{2}) (\uparrow\downarrow - \downarrow\uparrow)$.}
\end{figure}

\begin{figure}[ht]
\centerline{\includegraphics[width=0.85\linewidth,angle=0]{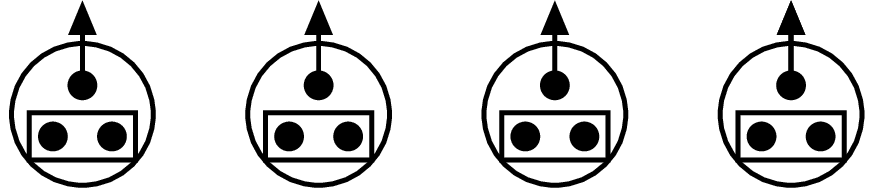}}%
\caption{\label{large-d} 
Schematic picture of the large-$D$ mechanism of the 1/3 plateau.
Two spins in a rectangle is in the state $(1/\sqrt{2}) (\uparrow\downarrow + \downarrow\uparrow)$.}
\end{figure}

\begin{figure}[ht]
\bigskip
\bigskip
\bigskip
\centerline{\includegraphics[width=0.85\linewidth,angle=0]{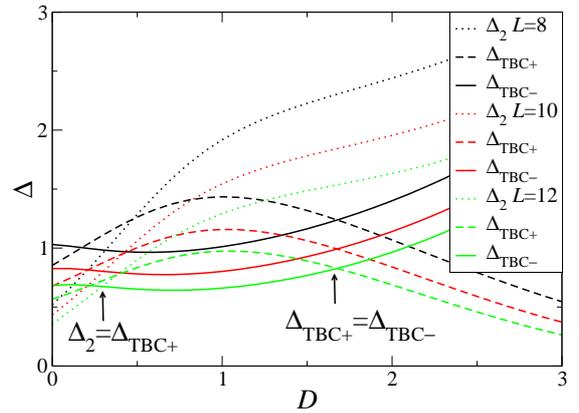}}%
\caption{\label{m1-3LS} 
(Color online) Three gaps $\Delta_1, \Delta_{\rm TBC+}, \Delta_{\rm TBC-}$ plotted versus $D$ 
with $\lambda$ fixed to 2.0 for $L=8, 10$ and 12. at $m=1/3$. 
}
\end{figure}

\begin{figure}[ht]
\bigskip
\bigskip
\centerline{\includegraphics[width=0.85\linewidth,angle=0]{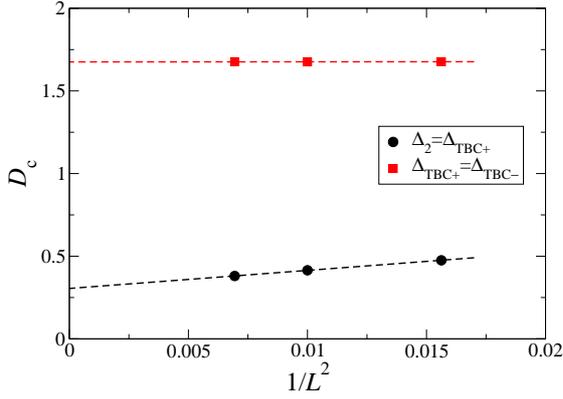}}%
\caption{\label{m1-3LSgaiso} 
(Color online) Estimation of the critical values of $D$ in the thermodynamic limit at $m=1/3$ when $\lambda = 2.0$.  
The estimated phase boundaries are $D_c=0.305\pm 0.002$ for the no-plateau and Haldane plateau phases,
and $D_c=1.675 \pm 0.001$ for the Haldane and large-$D$ ones.
}
\end{figure}

The two different magnetization plateaux for $Q=1$ at $m=1/3$ had been 
already predicted theoretically using the numerical diagonalization and the 
level spectroscopy analyses\cite{kitazawa}. 
One is the Haldane plateau shown in Fig. \ref{haldane} and the other 
is the large-$D$ plateau shown in Fig. \ref{large-d}. 
Figures \ref{haldane} and \ref{large-d} describe schematic pictures of 
the mechanism of the plateau, considering the 3/2 spin as the 
composite spin of three 1/2 spins. 
In order to extend the phase diagram to wider region of the 
anisotropy parameters $\lambda$ and $D$, 
we use the same method as Ref. 4), namely the level spectroscopy 
analysis.\cite{kitazawa,nomura-kitazawa}
Then we review this method briefly here. 
To distinguish these plateau phases and the no-plateau phase 
based on this method, 
we should compare the following three excitation gaps at $m=1/3$:  
\begin{eqnarray}
\Delta_2 &=&{E(L,M+2)+E(L,M-2)-2E(L,M) \over 2}, 
\\ \label{delta2}
\Delta_{\rm TBC+}&=&E_{\rm TBC+}(L,M)-E(L,M), 
\\ \label{tbcp}
\Delta_{\rm TBC-}&=&E_{\rm TBC-}(L,M)-E(L,M), 
\label{tbcm}
\end{eqnarray}
where $E_{\rm TBC+}(L,M)$ and $E_{\rm TBC-}(L,M)$ are the lowest energy 
eigenvalues of the even-parity and odd-parity wave functions with respect to 
the space inversion at the twisted boundary, respectively, and $M=L/2$. 
According to the level spectroscopy analysis, 
the smallest excitation gaps among them determine the phase at $m=1/3$. 
If $\Delta_2$ is the smallest, the system has no 1/3 plateau. 
If $\Delta_{\rm TBC+}$ ($\Delta_{\rm TBC-}$) is the smallest, 
the system is in the large-$D$ (Haldane) plateau phase. 
Fixing $\lambda$ to 2.0, 
the excitation gaps $\Delta_2$, $\Delta_{\rm TBC+}$ and $\Delta_{\rm TBC-}$
are plotted versus $D$ for $L$=8, 10 and 12
in Fig. \ref{m1-3LS}. It indicates that as $D$ increases, the smallest
gap changes from $\Delta_2$, through $\Delta_{\rm TBC-}$,
to $\Delta_{\rm TBC+}$. Thus the system has no plateau for small $D$,
the Haldane plateau for intermediate $D$, and the large-$D$
one for large $D$. 
Assuming the system size correction being proportional to $
1/L^2$,
The cross point between $\Delta_2$ and $\Delta_{\rm TBC-}$, and that
between $\Delta_{\rm TBC-}$ and $\Delta_{\rm TBC+}$
are extrapolated to the infinite $L$ limit 
as shown in Fig. \ref{m1-3LSgaiso}.
These procedures result in the estimated phase boundaries as $D_c=0.305
\pm 0.002$ for the no-plateau and Haldane plateau phases,
and $D_c=1.675 \pm 0.001$ for the Haldane and large-$D$ ones.

\begin{figure}[ht]
\centerline{\includegraphics[width=0.85\linewidth,angle=0]{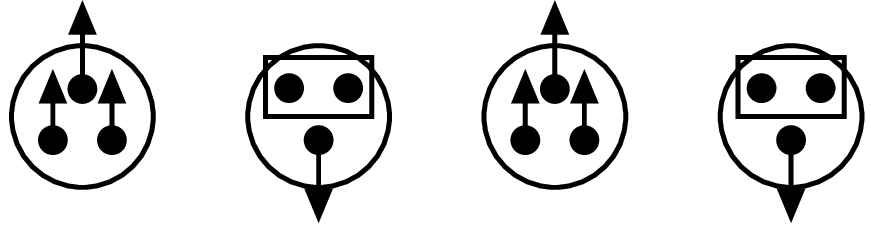}}%
\caption{\label{neel} 
Schematic picture of the Neel mechanism of the 1/3 plateau.
}
\end{figure}

\begin{figure}[ht]
\bigskip
\bigskip
\centerline{\includegraphics[width=0.85\linewidth,angle=0]{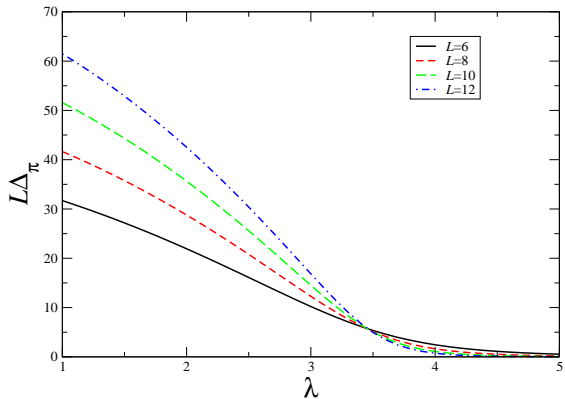}}%
\caption{\label{m1-3prg} 
(Color online) Scaled gaps $L\Delta_{\pi}$ plotted versus $\lambda$ for $L=6, 8, 10$ and 12 
at $m=1/3$ when $D=2.0$. 
}
\end{figure}

\begin{figure}[ht]
\bigskip
\bigskip
\bigskip
\centerline{\includegraphics[width=0.85\linewidth,angle=0]{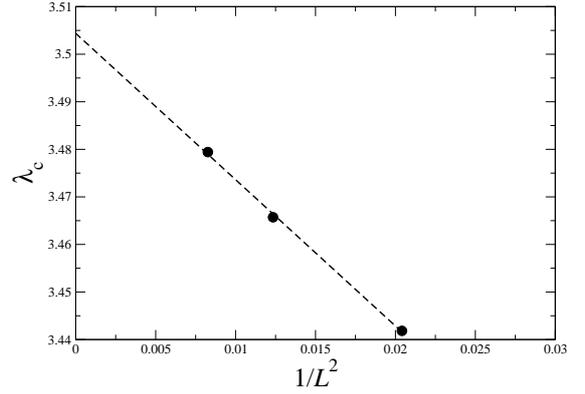}}%
\caption{\label{m1-3prg-gaiso} 
The size-dependent fixed point $\lambda_c$ is plotted versus $1/L^2$ for $D=2.0$. 
The estimated critical point is $\lambda_c =3.504 \pm 0.001$. 

}
\end{figure}

Next we consider the translational-symmetry-broken plateau for $Q=2$ at $m=1/3$. 
It is expected to be the N\'eel plateau like 
$|\cdots {3\over 2},{-{1\over 2}},{3\over 2},{-{1\over 2}},
{3\over 2},{-{1\over 2}},\cdots \rangle$ in the large $\lambda$ limit. 
The schematic picture of the mechanism is shown in Fig. \ref{neel}. 
The phenomenological renormalization\cite{PRG} is a good method to determine 
the phase boundary between the $Q=1$ and $Q=2$ plateau phases. 
We apply this method to the excitation gap
\begin{equation}
\Delta_{\pi}(L,\lambda,D)=E_{k=\pi}(L,M)-E(L,M),
\label{pi}
\end{equation}
where $E_{k=\pi}(L,M)$ is the lowest energy eigenvalue in the subspace 
for $k=\pi$ and $M=L/2$. 
Let us consider the scaled gap $L\Delta_{\pi}(L,\lambda,D)$.
In the $Q=2$ plateau region (namely, the N\'eel plateau region),
the plateau state is two-fold degenerate in the thermodynamical limit.
In the finite system size case,
the low-lying excited state with $E_{k=\pi}(L,M)$ gradually degenerate to the
ground state with $E(L,M)$ along the way $\Delta_{\pi}(L,\lambda,D) \sim  \exp(-aL)$
as $L \to \infty$, where $a$ is a positive constant.
Thus the scaled gap behaves as $L\Delta_{\pi}(L,\lambda,D) \sim  L\exp(-aL)$,
which is a decreasing function of $L$.
On the other hand, in the $Q=1$ plateau region,
since $\Delta_{\pi}(L,\lambda,D)$ has a finite value in the $L \to \infty$ limit,
the scaled gap behaves as $L\Delta_{\pi}(L,\lambda,D) \sim L$.
Furthermore, on the $Q=1$ and $Q=2$ boundary line,
which is expected in the Ising universality class,
the behavior of the excitation gap will be $\Delta_{\pi}(L,\lambda,D) \sim 1/L$,
which leads to $L\Delta_{\pi}(L,\lambda,D) \sim {\rm const}$.
According to the above consideration,
the size-dependent critical point $\lambda_c$ 
is derived from the phenomenological renormalization  fixed point equation
\begin{eqnarray}
L\Delta_{\pi}(L,\lambda_c,D)=(L+2)\Delta_{\pi}(L+2,\lambda_c,D),
\label{prg}
\end{eqnarray}
for each value of $D$. 
When $D$ is fixed to 2.0, the scaled gap $L\Delta_{\pi}$ is plotted versus $\lambda$ 
for $L=6, 8, 10$ and 12 in Fig. \ref {m1-3prg}. 
Assuming that the size correction of $\lambda_c$ determined for $L$ and $L+2$ 
is proportional to $1/(L+1)^2$, 
the critical point $\lambda_c$ in the infinite $L$ limit is estimated 
as shown in Fig. \ref{m1-3prg-gaiso}. 
The result is $\lambda_c =3.504 \pm 0.001$.

\begin{figure}[ht]
\bigskip
\bigskip
\centerline{\includegraphics[width=0.85\linewidth,angle=0]{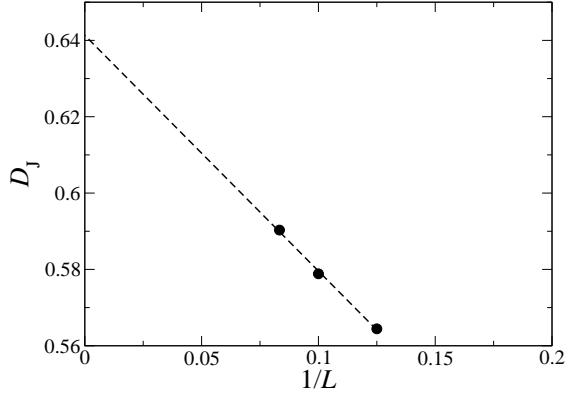}}%
\caption{\label{jump} 
Points for $R(L,M)=0$ are plotted versus $1/L$ for $\lambda=4.0$. 
Assuming the size correction proportional to $1/L$, 
the estimated critical value $D_J$ in the infinite $L$ limit is $0.641 \pm 0.003$. 
}
\end{figure}

It is also found that there is a region where the magnetization jump occurs and the $m=1/3$ state is not realized 
for sufficiently large $\lambda$ and small $D$. 
When the lowest energy eigenvalue per unit cell is defined as $\epsilon (m)$ for the magnetization $m$, 
the condition for the skip of the magnetization $m$ is $\epsilon''(m) < 0$\cite{sakai6}. 
If we define $R(L,M)$ as the form
\begin{eqnarray}
R(L,M) \equiv L[E(L,M+1)+E(L,M-1)-2E(L,M)], 
\label{rdef}
\end{eqnarray}
it satisfies the relation
\begin{eqnarray}
R(L,M) \rightarrow \epsilon''(m) \quad (L\rightarrow \infty). 
\label{rinf}
\end{eqnarray}
Thus the boundary of the region where $m=1/3$ is skipped can be estimated as the points for $R(L,M)=0$ in the infinite
$L$ limit. 
We estimate these points for $L$=8, 10 and 12, and extrapolate them to the infinite $L$ limit,
assuming the size correction being proportional to $1/L$. 
For example, when $\Delta$ is fixed to 4.0, the points for $R(L,M)=0$ are plotted versus $1/L$ in 
Fig. \ref{jump}. The estimated critical value $D_J$ is $0.641 \pm 0.003$. 

\begin{figure}[ht]
\bigskip
\bigskip
\bigskip
\centerline{\includegraphics[width=0.85\linewidth,angle=0]{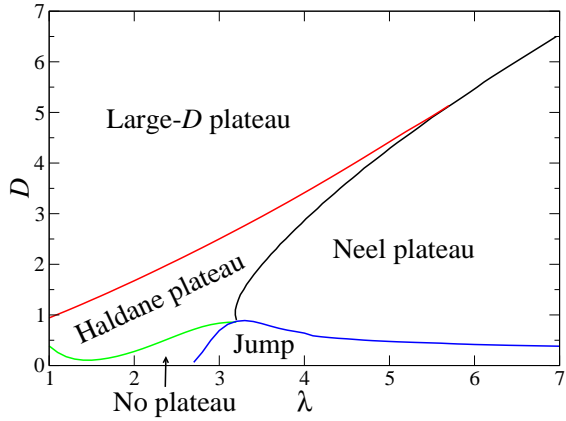}}%
\caption{\label{m1-3phase} 
(Color online) Phase diagram at $m$=1/3. 
`Jump' means the region where $m=1/3$ is not realized because it is skipped by the magnetization jump. 
}
\end{figure}

The phase diagram with respect to the anisotropies $\lambda$ and $D$ at $m=1/3$ is 
obtained as Fig. \ref{m1-3phase}. 
It includes wider region of $\lambda$ and $D$ than the previous work\cite{kitazawa}
where the region $0 < \lambda < 1$ was discussed. 
Then the N\'eel plateau phase and the jump region where $m=1/3$ is skipped is found
for the first time. 

\subsection{$m=2/3$}

\begin{figure}[ht]
\centerline{\includegraphics[width=0.85\linewidth,angle=0]{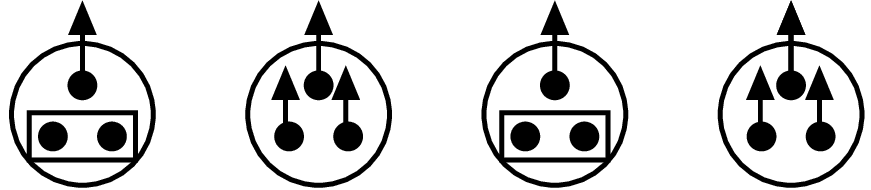}}%
\caption{\label{2-3neel} 
Schematic picture of the Neel mechanism of the 2/3 plateau.
}
\end{figure}

The possibility of the $m = 2/3$ magnetization plateau is investigated. 
Since $Q=2$ is necessary, the N\'eel plateau like 
$|\cdots, {3\over 2},{{1\over 2}},{3\over 2},{{1\over 2}},
{3\over 2},{{1\over 2}},\cdots \rangle$ is expected to appear. 
The schematic picture of it is shown in Fig. \ref{2-3neel}. 

\begin{figure}[ht]
\bigskip
\bigskip
\centerline{\includegraphics[width=0.85\linewidth,angle=0]{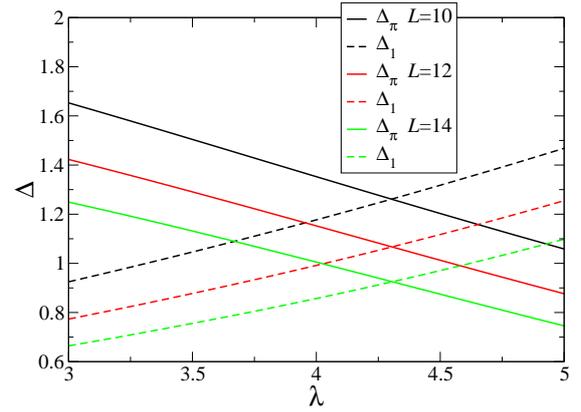}}%
\caption{\label{m2-3LS} 
(Color online) Gaps $\Delta_{\pi}$ and $\Delta_1$ plotted versus $\lambda$ with $D$ fixed to 5.0 
for $L=$10, 12 and 14. 
}
\end{figure}

\begin{figure}[ht]
\bigskip
\bigskip
\bigskip
\centerline{\includegraphics[width=0.85\linewidth,angle=0]{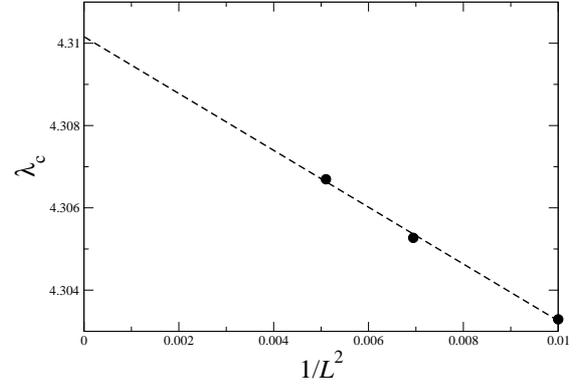}}%
\caption{\label{m2-3LSgaiso} 
Estimation of the critical values of $\lambda$ for $D$=5.0 in the thermodynamic limit at $m=2/3$, 
assuming the size correction proportional to $1/L^2$. 
The result is $\lambda_c=4.310 \pm 0.001$. 
}
\end{figure}

In the N\'eel plateau phase, the ground state should be 
doubly degenerate and the energy gap would be open. 
In order to determine the boundary between the 
N\'eel-plateau and no-plateau phases, 
another level spectroscopy analysis\cite{okamoto-nomura,nomura-okamoto, okamoto2002}
different from the previous subsection 
is useful. 
In this method, we should compare the following two excitation gaps:
\begin{eqnarray}
\Delta_1 &=& {E(L,M+1)+E(L,M-1)-2E(L,M) \over 2},  
\label{1mgap} \\
\Delta_{\pi} &=& E_{k=\pi}(L,M)-E(L,M), 
\label{pigap} 
\end{eqnarray}
at $m=2/3$, namely $M=L$. 
$\Delta_{\pi}$ is the same as Eq.(\ref{pi}). 
If $\Delta_1$ ($\Delta_{\pi}$) is the smaller, the system is in the no-plateau (N\'eel plateau) phase. 
When $D$ is fixed to 5.0, these gaps are plotted versus $\lambda$ for $L=10, 12$ and 14 
in Fig. \ref{m2-3LS}. 
Assuming that the size correction of the cross points between them is proportional to $1/L^2$, 
the phase boundary $\lambda_c$ in the thermodynamic limit is estimated as shown in Fig. \ref{m2-3LSgaiso}.

\begin{figure}[ht]
\bigskip
\bigskip
\centerline{\includegraphics[width=0.85\linewidth,angle=0]{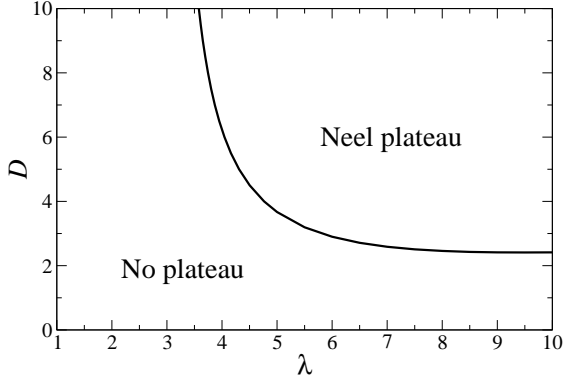}}%
\caption{\label{m2-3phase} 
Phase diagram at $m=2/3$.
}
\end{figure}

Using this method, the phase diagram at $m=2/3$ is obtained as shown in Fig. \ref{m2-3phase}. 
The shape of the phase diagram is quite different from that of $m=1/3$.
Namely, only the N\'eel plateau phase appears.  
We note that in the very large $\lambda$ case beyond Fig. \ref{m2-3phase} ($\lambda \gtrsim 10.3$),
the jump region also appears in the phase diagram at $m=2/3$.

\section{Magnetization Curves}

\begin{figure}[ht]
\bigskip
\bigskip
\centerline{\includegraphics[width=1.0\linewidth,angle=0]{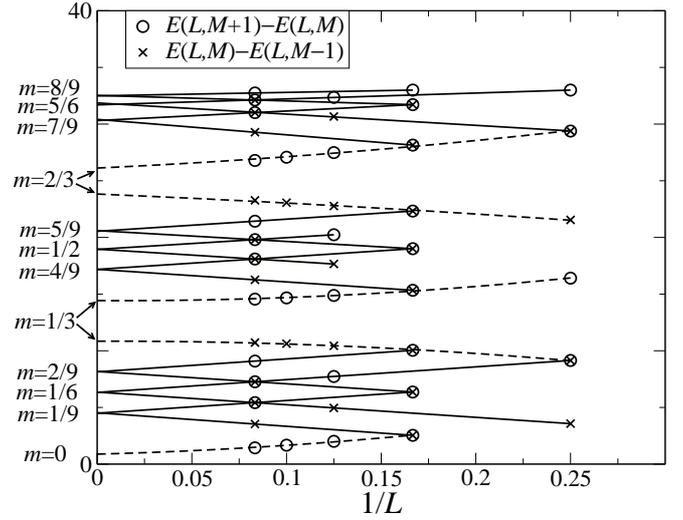}}%
\caption{\label{extramag} 
Magnetization curves for $\lambda=6.0$ and $D=6.0$. 
}
\end{figure}

\begin{figure}[ht]
\bigskip
\bigskip
\bigskip
\bigskip
\centerline{\includegraphics[width=0.95\linewidth,angle=0]{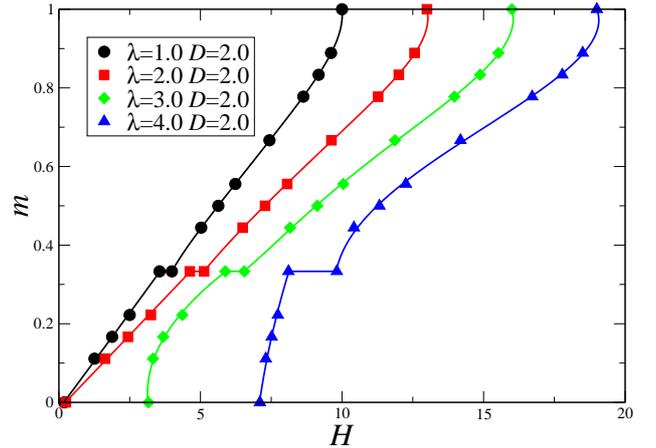}}%
\caption{\label{magd20} 
(Color online) Magnetization curves for the following parameters: 
(1) $(\lambda,D)=(1.0, 2.0)$; 1/3 large-$D$ plateau, (2) $(\lambda, D)=(2.0, 2.0)$; 
1/3 large-D plateau, (3) $(\lambda,D)=(3.0,2.0)$; 1/3 Haldane plateau, 
(4) $(\lambda,D)=(4.0, 2.0)$; 1/3 N\'eel plateau. 
The analytical expression for the saturation field, Eq.(\ref{eq:saturation}),
gives $10,13,16$ and 19 for $\lambda = 1.0,2.0,3.0$ and 4.0, respectively.}
\end{figure}

\begin{figure}[ht]
\bigskip
\bigskip
\bigskip
\bigskip
\centerline{\includegraphics[width=0.95\linewidth,angle=0]{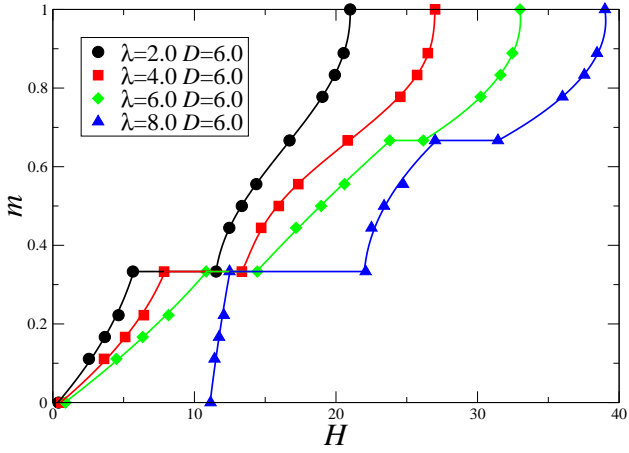}}%
\caption{\label{magd60} 
(Color online) Magnetization curves for following parameters: 
(1) $(\lambda,D)=(2.0, 6.0)$; 1/3 large-$D$ plateau, 
(2) $(\lambda,D)=(4.0, 6.0)$; 1/3 large-$D$ plateau, 
(3) $(\lambda,D)=(6.0, 6.0)$; 1/3 Large-$D$ and 2/3 N\'eel plateaux, 
(4) $(\lambda,D)=(8.0, 6.0)$; 1/3 and 2/3 N\'eel plateaux. 
The analytical expression for the saturation field, Eq.(\ref{eq:saturation}),
gives $21,27,33$ and 39 for $\lambda = 2.0,4.0,6.0$ and 8.0, respectively.}
\end{figure}

In order to encourage the experimental study to discover the magnetization plateau, 
we calculate the ground-state magnetization curves for several typical parameters. 
When the system is of no-plateau at $m$, $E(L,M+1)-E(L,M)$ and $E(L,M)-E(L,M-1)$ for $M={3\over 2}Lm$ 
have the asymptotic forms in the infinite $L$ limit, 
\begin{eqnarray}
E(L,M+1)-E(L,M) &\sim& H(m) +O(1/L),
\label{h2} \\
E(L,M)-E(L,M-1) &\sim& H(m) + O(1/L),
\label{h1} 
\end{eqnarray}
where $H(m)$ is the magnetic field for $m$ in the infinite $L$ limit. 
These quantities are plotted versus $1/L$ for $\lambda=6.0$ and $D=6.0$ 
in Fig. \ref{extramag}. 
It justifies the relations (\ref{h2}) and (\ref{h1}) 
except for the plateau cases; $m=0, 1/3$ and 2/3. 
In the no-plateau cases, we estimate $H(m)$ assuming the following asymptotic form:
\begin{eqnarray}
[E(L,M+1)-E(M-1)]/2  \sim H(m) +O(1/L^2). 
\label{field}
\end{eqnarray}
On the other hand, in the plateau cases, 
we use the Shanks transformation defined as the form
\begin{eqnarray}
P_L'={{P_{L-2}P_{L+2}-P_L^2}\over{P_{L-2}+P_{L+2}-2P_L}}.
\label{shanks}
\end{eqnarray}
We apply this transformation to $P_L = E(L,M+1)-E(L,M)$ with $L=10$ to 
estimate the higher edges of the plateaux at $m=0, 1/3$ and 2/3, 
and to $P(L) = E(L,M)-E(L,M-1)$ to estimate the lower edges of the plateaux 
at $m=1/3$ and 2/3. 
Using this method, 
we obtain the magnetization curves for several typical anisotropy parameters, 
shown in Figs. \ref{magd20} and \ref{magd60}. 
In Fig. \ref{magd20} for $D=2.0$, the magnetization curve has the 1/3 large-$D$ plateau for $\lambda =1.0$ and $\lambda=2.0$, 
the 1/3 Haldane one for $\lambda=3.0$, and the 1/3 N\'eel one for $\lambda=4.0$. 
In Fig. \ref{magd60} for $D=6.0$, it has the 1/3 large-$D$ plateau for $\lambda=2.0$ and 4.0, 
the 1/3 large-$D$ and  2/3 N\'eel ones for $\lambda=6.0$, and the 1/3 and 2/3 N\'eel ones for $\lambda=8.0$. 
In Figs. \ref{magd20} and \ref{magd60}, solid symbols are the estimated points in the infinite $L$ limit, 
and curves are guides for the eye.  

The saturation field $H_\rs$ can be analytically estimated by
calculating the energy difference between the ferromagnetic state and
the one-spin-down state.
We obtain
\begin{equation}
    H_\rs = 3\lambda + 2D + 3,
    \label{eq:saturation}
\end{equation}
which well explains the numerically calculated values 
as written in the captions of Figs. \ref{magd20} and \ref{magd60}.

\section{Discussion}

\begin{table}[ht]
  \caption{Summary of the plateaux of the present and related models.
  Here $S$, $m$, and $\tilde m$ are
  the magnitude of the spin, relative magnetization defined by Eq.(\ref{eq:rel-mag}),
  magnetization per unit cell, 
  and $Q$ and $n$ are the parameters of Eq.(\ref{condition}), respectively.
  There is a relation $Sm = \tilde m$.
  The Haldane plateau, the large-$D$ plateau and
  the N\'eel plateau are denoted by H, LD and N, respectively.}
  \label{table-1}
  \begin{tabular}{|c|c|c|c|c|c|c|c|c|} \hline
     $S$ 
     &$m$
     &${\tilde m}$ 
     &$S-{\tilde m}$
     &$Q$
     &$n$
     &plateau type
     &
     &Refs.
     \\ \hline
         &    &    &    &1 &1 &H, LD                    &(a)& \\ \cline{5-8}
     \raisebox{0.5\normalbaselineskip}[0pt][0pt]{1}
        &\raisebox{0.5\normalbaselineskip}[0pt][0pt]{0}
        &\raisebox{0.5\normalbaselineskip}[0pt][0pt]{0}
        &\raisebox{0.5\normalbaselineskip}[0pt][0pt]{1}
        &2 &2 &N                                        &(b)
        &\raisebox{0.1\normalbaselineskip}[0pt][0pt]{\normalsize \cite{den-nijs,chen-S=1}} \\ \hline
     1   &1/2 &1/2 &1/2 &2 &1 &N                        &(c)
        &\raisebox{-0.35\normalbaselineskip}[0pt][0pt]{\normalsize \cite{sakai4}} \\ \hline
         &    &    &    &1 &2 &see Refs.\cite{tone-S=2,kjall} &(d)& \\ \cline{5-8}
     \raisebox{0.5\normalbaselineskip}[0pt][0pt]{2}
        &\raisebox{0.5\normalbaselineskip}[0pt][0pt]{0}
        &\raisebox{0.5\normalbaselineskip}[0pt][0pt]{0}
        &\raisebox{0.5\normalbaselineskip}[0pt][0pt]{2}
         &2 &4 &see Refs.\cite{tone-S=2,kjall}                &(e)
         &\raisebox{0.1\normalbaselineskip}[0pt][0pt]{\normalsize \cite{tone-S=2,kjall}} \\ \hline
         &    &    &    &1 &1 &H, LD                    &(f)& \\ \cline{5-8}   
     \raisebox{0.5\normalbaselineskip}[0pt][0pt]{2}
        &\raisebox{0.5\normalbaselineskip}[0pt][0pt]{1/2}
        &\raisebox{0.5\normalbaselineskip}[0pt][0pt]{1}
        &\raisebox{0.5\normalbaselineskip}[0pt][0pt]{1}
         &2 &2 &N                                       &(g)
        &\raisebox{0.1\normalbaselineskip}[0pt][0pt]{\normalsize \cite{sakai2019,yamada}} \\ \hline
         &    &    &    &1 &1 &H, LD                    &(h)
         &\raisebox{-0.5\normalbaselineskip}[0pt][0pt]{\normalsize \cite{kitazawa}} \\ \cline{5-8}
     \raisebox{0.5\normalbaselineskip}[0pt][0pt]{3/2}
        &\raisebox{0.5\normalbaselineskip}[0pt][0pt]{1/3}
        &\raisebox{0.5\normalbaselineskip}[0pt][0pt]{1/2}
        &\raisebox{0.5\normalbaselineskip}[0pt][0pt]{1}
           &2 &2 &N                                     &(i)&present \\ \hline  
     3/2 &2/3 &1   &1/2 &2 &1 &N                        &(j)&present \\ \hline
  \end{tabular}
\end{table}

We have obtained the phase diagrams at $m=1/3$ and $m=2/3$.
We summarize the types of plateaux of the present and related models in Table \ref{table-1}.
Comparing the phase diagrams of these models,
our $(S,m)=(3/2,1/3)$ phase diagram (cases (h) and (i)) is similar to 
those of $(S,m)=(1,0)$ (cases (a) and (b))
 and $(S,m)=(2,1/2)$ (cases (f) and (g)).
In fact, there appears the Haldane plateau, the large-$D$ plateau and the N\'eel plateau
in these models,
and also their dispositions of the phases are similar to one another.
On the other hand, in our case (j) with $(S,m)=(3/2,2/3)$,
there appears only the N\'eel plateau phase,
which is similar to the case (c) with $(S,m) = (1,1/2)$.
We note that only the N\'eel spin gap appears in the model (\ref{ham}) with $S=1/2$
(the $D$ term is a constant)
for $m=0$.

The key point is the value of $S-\tilde m$.
This value is $S-\tilde m = 1$ for our $(S,m)=(3/2,1/3)$ case and similar cases,
while $S-\tilde m = 1/2$  for our $(S,m)=(3/2,2/3)$ case and similar cases. 
In the composite spin picture.
an $S=3/2$ spin is composed of three $s=1/2$ component spins,
as is shown in  Figs.\ref{haldane}, \ref{large-d} and \ref{neel}.
When $m=1/3$, one of three component spins turns to the $z$-direction to
maintain the magnetization
and the remaining two component $s=1/2$ spins are free to couple with another
component spin.
The number of free component spins per an $S$ spin is nothing but
$S - \tilde m$.
This situation is very similar to the $S=1$ chain with $m=0$.
Thus, referring the phase diagram of $(S,m) = (1,0)$,\cite{den-nijs,chen-S=1}
there appears the Haldane plateau, the large-$D$ plateau and
the N\'eel plateau in our $(S,m) = (3/2,1/3)$ phase diagram.
For our $m=2/3$ case,
two component $s=1/2$ spins turn to the $z$-direction,
resulting in only one free $s=1/2$ component spin per an $S$ spin.
This state of affairs is the same as that of $S=1/2$ XXZ chain,
which has the N\'eel state as the only gapped state.

The phase diagram of the $(S,m) = (2,0)$ case\cite{tone-S=2,kjall}
is quite different from those of other cases,
reflecting $S-\tilde m =2$.
For instance, there is no essential difference between the Haldane
phase and the large-$D$ phase.
If we apply the consideration based on $S-\tilde m$,
the $(S,m) = (5/2,1/5)$ and $(S,m) = (3,1/3)$ cases
are in the similar situation.
This will be a future problem.

The phase boundary between the large-$D$ plateau and N\'eel plateau
of Fig. \ref{m1-3phase}
in the $\lambda \gg 1$ and $D \gg 1$ can be explained
in the following way.
In this case the coupling of $S_j^x S_{j+1}^x + S_j^y S_{j+1}^y$
can be neglected.
Then, referring the schematic pictures of these plateau mechanism,
Figs. \ref{large-d} and \ref{neel},
these energies at $m=1/3$ are
\begin{eqnarray}
   &&E_{\rm LD} = {L(\lambda + D - 2H) \over 4},
   \label{eq:LD}\\
   &&E_{\rm Neel} = {L(-3\lambda + 5D - 2H) \over 4},
   \label{eq:Neel}
\end{eqnarray}
respectively.
Therefore the boundary between these two phases is
\begin{equation}
   D = \lambda, 
   \label{eq:boundary1}
\end{equation}
which well explains the phase diagram of Fig. \ref{m1-3phase}.

In the phase diagram of $m=2/3$, Fig. \ref{m2-3phase},
the $D>0$ situation is needed for the realization of the N\'eel plateau.
At a glance this seems to be curious
because $D>0$ is unfavorable to the N\'eel state as shown
in the phase diagram of $(S,m)=(1,0)$.\cite{den-nijs,chen-S=1}
The composite spin picture of the $m=2/3$ N\'eel plateau state is
shown in Fig. \ref{2-3neel},
where the $S^z=3/2$ and $S^z=1/2$ spins are arrayed alternatingly.
In the usual N\'eel case such as $S=1/2$ chain under zero magnetic field,
isolated spins with $S^z=\pm 1/2$ have same energies.
On the other hand, in the present plateau case,
the energies of isolated $S^z=3/2$ and $S^z=1/2$ are different from
each other due to the magnetic field.
Thus $D>0$ is needed to compensate this energy difference. 
According to our investigation, the magnetization plateau
does not appear in the opposite competing case 
(namely for $\lambda <1$ and $D<0$),
which is consistent with the above consideration..


The phase boundary between the N\'eel plateau phase and the no plateau phase
of Fig.\ref{m2-3phase} at $m=2/3$ can be explained by
an effective theory when $D \to \infty$.
In this case we can consider that half of the spins are in the $S^z=3/2$ state
and the remaining half are in the $S^z=1/2$ state
because other two states have much higher energies when $D \to \infty$.
We introduce the pseudo-spin operator with $T=1/2$ as
\begin{eqnarray}
   &&S_j^\pm = \sqrt{3}T^\pm
   \label{eq:Tpm}\\
   &&S_j^z = 1 + T_j^z
   \label{eq:Tz}
\end{eqnarray}
to pick up above two states.
Then we obtain the effective Hamiltonian as
\begin{eqnarray}
   &&\hskip-1cm {\cal H}_{\rm eff}
   =\sum_{j=1} ^L\left\{ 3(T_j^x T_{j+1}^x + T_j^y T_{j+1}^y) + \lambda T_j^z T_{j+1}^z \right\} \nonumber \\
     &&+ (2\lambda + 2D - H) \sum_{j=1}^L T_j^z + L\left(\lambda + {5D \over 4} - H\right),
   \label{eq:heff}
\end{eqnarray}
where $m=2/3$ of the original system corresponds to the zero magnetization of the effective system
described by $T$.
Then the magnetic field of $m=2/3$ is
\begin{equation}
   H_{2/3} = 2\lambda + 2D.
   \label{eq:h2-3}
\end{equation} 
When there is a plateau at $m=2/3$, this $H_{2/3}$ is the field of the center of the plateau.
The ${\cal H}_{\rm eff}$ system exhibits the transition
between the Tomonaga-Luttinger liquid state and the N\'eel state when $H=H_{2/3}$
at $\lambda = 3$ \cite{giamarchi},
which corresponds to the transition between the no-plateau state and the N\'eel plateau
state of the original system.
Thus the behavior of the phase boundary of Fig. \ref{m2-3phase},
$\lambda \to 3$ as $D \to \infty$,
is semi-quantitatively explained.
The magnetic field of $m=2/3$ (center of the plateau when plateauful)
is also well explained by Eq.(\ref{eq:h2-3})
as can be seen in Fig. \ref{magd60} for $D=6.0$.
In fact, from  Eq.(\ref{eq:h2-3}),
we see $H_{2/3} = 16,~20,~24$ and $28$ for 
$\lambda =2.0,~4.0,~6.0$ and $8.0$ respectively.


\section{Summary}

The magnetization process of the $S=3/2 $ antiferromagnetic chain with the exchange and the single-ion anisotropies 
is investigated using the numerical diagonalization of finite-size clusters and some size scaling analyses. 
In the case of the competing anisotropies, namely the easy-axis coupling anisotropy ($\lambda >1$) and 
the easy-plane single-ion one ($D>0$), 
the translational-symmetry-broken magnetization plateaux are revealed to appear at $m=1/3$ and $m=2/3$ 
for sufficiently large anisotropies. 
The phase diagram at $m=1/3$ including the $Q=1$ and $Q=2$ plateau phases, and the region 
where $m=1/3$ is skipped because of the magnetization jump is presented. 
The phase diagram at $m=2/3$ consisting of the no-plateau and plateau phases is also obtained. 
In the $m=2/3$ case $D>0$ favors the N\'eel plateau,
of which reason is physically explained.,
Some characteristics of the phase diagrams can be explained analytically.
In addition the magnetization curves for several typical parameters are shown. 
We hope some candidate materials suitable for such interesting magnetization curves will be discovered 
in the near future.

\section*{Acknowledgment}
This work was partly supported by JSPS KAKENHI, 
Grant Numbers JP16K05419, JP20K03866, JP16H01080 (J-Physics), 
JP18H04330 (J-Physics), JP20H05274 and 23K11125.
A part of the computations was performed using
facilities of the Supercomputer Center,
Institute for Solid State Physics, University of Tokyo,
and the Computer Room, Yukawa Institute for Theoretical Physics,
Kyoto University.
We used the computational resources of the supercomputer 
Fugaku provided by the RIKEN through the HPCI System 
Research projects (Project ID: hp200173, hp210068, hp210127, 
hp210201, hp220043, hp230114, hp230532, and hp230537
). 


\end{document}